\documentstyle[aasms4,11pt,epsf]{article}

\begin{document}

\title{The [OIII] Emission-line Nebula of the $z=3.594$ Radio Galaxy 
4C $+$19.71$^{*}$}
\author{L. Armus$^1$, B.T. Soifer, T.W. Murphy Jr., G. Neugebauer, A.S. Evans 
\& K. Matthews}
\vskip 0.5truein
\affil{Palomar Observatory, Caltech, Pasadena, CA 91125}
\affil{$^1$ Current Address, Infrared Processing and Analysis Center, Caltech 100-22, Pasadena, CA 91125}
\affil{$^*$ Based on observations at the W.M. Keck Observatory, which is 
operated by the California Institute of Technology and the University 
of California}

\begin{abstract} 

We have imaged the $z=3.594$ radio galaxy 4C $+$19.71 in the light of
the redshifted [OIII] 5007\AA~ emission line, using a narrow-band
filter centered at 2.3$\mu$m with the Near Infrared Camera on the
Keck Telescope.  The [OIII] nebula of 4C $+$19.71 has a size of
$74\times9$ kpc, and a luminosity of L$_{5007}\sim3 \times10^{37}$
W.  The rest frame equivalent width of the 5007\AA~ line, averaged
over the entire nebula, is  560\AA.  The length of the major axis of
the [OIII] emission is nearly identical to the separation of the
radio lobes seen at 1465 MHz (Rottgering, et al. 1994), and the
position angle of the nebula is the same as that of the two radio
lobes.  In addition, 4C $+$19.71 follows the optical emission line
vs. radio power correlation seen in other powerful radio galaxies.
The [OIII] and Ly$\alpha$ emission-line luminosities suggest that the
ionized gas mass lies in the range of
$2\times10^{8}-10^{9}$M$_{\odot}$.  The O/H ratio in the nebula is at
least a few tenths solar, and may be as high as a factor of three
above solar, indicating a previous phase of star formation in 4C
$+$19.71.  Thirty five percent of the total K-band flux is
contributed by the 5007\AA ~emission line, and the continuum of 4C
$+$19.71 has a K$\sim19.6$ mag.  This places 4C $+$19.71 along the
K-z relation found for other radio galaxies and radio loud quasars.
If the continuum is dominated by starlight, the host galaxy has a
rest frame visual luminosity of about 40L$^{\ast}$.  There are no
candidate emission-line objects at the redshift of 4C $+$19.71 having
[OIII] rest frame equivalent widths of more than about 2\% that of
the radio galaxy itself within a volume of 212 Mpc$^{3}$.

\end{abstract}


\section{Introduction}

A ubiquitous feature in the spectra of powerful radio galaxies is the
presence of bright ultraviolet and optical emission lines.  These
lines can be quite strong, with rest frame equivalent widths of
several hundred angstroms.  The strength of the emission lines,
coupled with the large sky coverage of radio surveys, has facilitated
the discovery of radio galaxies over a large range in redshift.  As a
result, powerful radio galaxies are among the most extensively
studied stellar systems at moderate and high redshifts.

By observing the emission-line properties of radio galaxies at
different redshifts, we can study the interplay between active
galactic nuclei and their host galaxies as a function of cosmic
epoch.  The emission-line nebulae provide both a direct probe of the
conditions in the interstellar medium of the host, and an indirect
probe of the central source of ionizing radiation.  The most
spectacular emission-line nebulae in these systems can have
luminosities above $10^{37}$ W, sizes of over 100 kpc, and a strong
alignment with the radio axis (McCarthy et al. 1987, Baum et al.
1988, Chambers, Miley \& van Breugel 1987,1988, McCarthy \& van
Breugel 1989, and McCarthy et al. 1995).

The emission line nebulae of high redshift radio galaxies are usually
studied via the redshifted Ly$\alpha$ line, which can, in principle,
be strongly affected by extinction from even a small amount of dust
due to resonant scattering.  Although most powerful radio galaxies
are bright in Ly$\alpha$ (McCarthy \& Lawrence 1993), dust and
density stratification in the interstellar gas can have a strong
effect on the flux and spatial distribution of the line-emitting
material seen in the rest frame ultraviolet.  The distorted
morphologies (Heckman et al.  1986), large infrared luminosities
(Golombek, Miley \& Neugebauer 1988), and molecular gas contents
(Mirabel, Sanders \& Kazes 1989; Mazzarella et al. 1993; Evans 1996)
of radio galaxies at low-redshift imply a great deal of dust in these
systems.  It might be expected therefore, that these conditions exist
in some high-redshift radio galaxies as well.  For example, 4C
$+$41.17 and 8C1435$+$63 have been found to have sub--mm flux
densities consistent with thermal dust emission (Dunlop et al. 1994,
Ivison 1996).  Also, recent observations of 4C $+$05.41 (Dey, Spinrad
\& Dickinson 1995) and TX0211$-$122 (van Ojik et al. 1994) have
provided data in support of models of dusty systems at high
redshift.  If high-redshift radio galaxies are dusty like many of
their low redshift counterparts, the nebular properties (e.g. sizes,
morphologies and luminosities) gleaned from observations of
Ly$\alpha$ might be misleading.  Besides dust, associated HI
absorption systems can apparently have a strong effect on the
Ly$\alpha$ properties of high redshift galaxies (van Ojik et al.
1997) as well.  However, by combining Ly$\alpha$ and H$\alpha$ and/or
[OIII] images of high redshift radio galaxies it might be possible to
construct extinction and/or ionization ``maps" of the emission-line
nebulae.  Since much work has been devoted to the study of the
optical emission-line nebulae of low redshift galaxies, near infrared
images of radio galaxies at $z>2$ made through narrow-band filters
provide a means for direct comparison in the same rest-frame emission
features.

Besides being valuable laboratories in which to study the interplay
between relativistic radio plasmas and dense, galactic gas, the stellar
properties of high-redshift radio galaxies can offer significant leverage
on cosmological parameters.  Powerful radio sources at $z > 2-3$ are
likely to reside within massive galaxies which may still be growing via
accretion at a time when the Universe was only $10\%-20$\% of its current
age.  Thus, the host galaxies of high redshift AGN may provide a glimpse
of the formation and build up of the most massive stellar systems seen
around us today.  Large galaxies at high redshift may also mark the
locations of young clusters whose cannibalised members form the stellar
building blocks of the central host, and provide gas to stoke the engine
of the radio source.  

The combination of narrow-band filters, sensitive, large format
infrared arrays, and large aperture telescopes naturally lends itself
to searches for high redshift active, or star forming galaxies in
young clusters (see Bunker et al.  1995, Mannucci \& Beckwith 1995,
and Thompson, Mannucci \& Beckwith 1996).  While clustering around
powerful radio galaxies at $z\sim0.5$ seems to be enhanced compared
to lower redshifts (Hill \& Lilly 1991),  the clustering properties
of radio galaxies at $z>1$ are largely unknown, with a pair of
notable exceptions.  Narrow-band Ly$\alpha$ imaging and follow-up
spectroscopy of the field around the $z=3.14$ radio galaxy
MRC0316-257 confirm the existence of two faint galaxies at the same
redshift as the radio source (Le Fevre, et al. 1997).  In addition,
Pascarelle et al. (1996) have discovered $10-20$ objects in the field
of the $z=2.39$ radio galaxy 53W002 via HST Ly$\alpha$ imaging and
follow-up spectroscopy.  Thus at least two groups or clusters are
known around powerful radio galaxies at $z>2$.

As part of a program to study the continuum and emission-line
properties of high redshift AGN in the near-infrared, we have imaged
the $z=3.594$ radio galaxy, 4C $+$19.71 (MG 2141+19) using the W.M.
Keck Telescope.  4C $+$19.71 is a double-lobed, steep-spectrum, FRII
(Fanaroff and Riley class II - Fanaroff \& Riley 1974) radio source
(Rottgering et al. 1994).  The two radio lobes are separated by
8.1$''$ at a position angle on the sky of $176^{\rm \circ}$.  The
integrated flux density in the lobes at 1465 MHz is S$_{1465} \sim 3
\times 10^{-27}$ W m$^{-2}$ Hz$^{-1}$, implying an emitted
monochromatic power at a rest frequency of 6.73 GHz of approximately
$10^{28}$ W Hz$^{-1}$.  The galaxy associated with 4C $+$19.71 has
been imaged previously in the K-band by Eales \& Rawlings (1996) and
was shown to have two faint components separated by approximately
4$''$.  The redshift of the galaxy is $z=3.594$ (Spinrad et al.
1993).  Eales \& Rawlings present a K-band spectrum showing a weak
feature at $\sim2.3\mu$m, identified as the [OIII] 5007\AA ~line,
having a flux of $2.2\pm0.4\times10^{-18}$ W m$^{-2}$.  Here, we
present both broad-band K and narrow-band $2.3\mu$m images of 4C
$+$19.71, which reveal a resolved central component at K, and a
large, asymmetric emission line nebula aligned along the radio axis.
The total [OIII] 5007\AA ~ emission-line luminosity of the nebula is
about $3\times10^{37}$ W, and the linear extent along the major axis
is 74 kpc.

In the following sections we describe the observations and present
the imaging results.  In section 4 we relate the nebular properties
of 4C $+$19.71 to other powerful radio galaxies, estimate the rest
frame blue continuum luminosity of 4C $+$19.71, and discuss the
limits on clustering around the radio galaxy probed by our
narrow-band imaging.

Throughout this paper we adopt H$_{\rm o}=75$ km s$^{-1}$ Mpc$^{-1}$ and
q$_{\rm o}=0$, so that at the redshift of 4C $+$19.71, 
9.3 kpc projects to 1$''$ on the sky.

\section{Observations and Data Reduction} 

Observations of 4C $+$19.71 were made at the W.M. Keck Observatory on
the night of 5 September 1996 with the Near Infrared Camera (Matthews
\& Soifer 1994).  The radio galaxy was imaged through a standard
K-band, 2.0-2.45$\mu$m, filter as well as through a narrow,
2.284-2.311$\mu$m, filter designed to sample the CO stellar
absorption feature in zero redshift galaxies.  At a redshift of
$z=3.594$, the [OIII] 5007\AA~ emission-line feature, but not the
4959\AA~ feature, falls in the bandpass of the narrow-band filter.
The plate scale of the 256x256 InSb array is 0.15$''$ per pixel.  The
total integration time through the K filter was 1080 seconds while
the total integration time through the narrow band filter was 2700
seconds.  In each case, individual images of 60 and 150 seconds
duration, respectively, were taken with the galaxy moved by about
10$''$ on the array between successive exposures.  An offset guider
employing a visual wavelength CCD was used to guide the telescope.  A
nearby star (referred to as star A in Fig. 1) was visible in each
exposure and used to register the individual frames.  The conditions
were photometric during the observations, and UKIRT faint standard
stars (Casali \& Hawarden 1992) provided the flux calibration.  The
seeing during the observations was $0.4''-0.5''$ full width at half
maximum (FWHM).  To remove time-variable fluctuations in
illumination, separate sky and normalized flat-field frames were
created from the data for each three images, by taking the median of
the nearest 7-9 frames.  After being trimmed to a size of $251 \times
251$ pixels, the individual data frame are thus sky subtracted and
flat fielded and are shifted to a common dc level after known bad
pixels are flagged.  These processed images are then aligned, using
integer pixel shifts, and combined using a clipped mean algorithm.

\section{Results}

Near infrared K-band images of 4C $+$19.71 are presented in Figs. 1 and
2.  Fig. 1 is an image of the $49'' \times 49''$ field surrounding the
radio galaxy.  The FWHM of a stellar point source in Fig. 1 is 3.3 pixels,
or $0.5''$.  All confidently detected sources, both resolved and
unresolved, are labelled in Fig. 1.  The K-band magnitudes of these
sources are given in Table 1.  We estimate a point source ($3\sigma$)
detection limit of K$\sim22.5$ mag in the final mosaic.  

The radio galaxy 4C $+$19.71 has two obvious 2.2$\mu$m components
(see Fig 2 for an expanded picture) labelled as ``a" and ``b" after
Eales \& Rawlings (1996).  Both components are resolved in our image,
and component ``a" has a faint extension to the south.  The
magnitudes of components ``a" and ``b", as measured through $2.0''$
diameter circular apertures, are K$=20.14\pm0.05$ mag and
K$=21.41\pm0.16$ mag, respectively.  The total K magnitude of 4C
$+$19.71 as measured from our data through a rectangular aperture of
$2''\times10''$, oriented north-south, centered on component ``a",
and excluding G6, is K$=18.92\pm0.04$ mag.  In Table 1, the K-band
magnitudes of all the sources identified in Figure 1 are given, as
are the positions relative to component ``a" of 4C $+$19.71.

In Fig. 2 we show the central $\sim20''$ region of the 4C $+$19.71
field imaged through both the broad-band K and narrow-band $2.3\mu$m
filters.  The difference in the appearance of 4C $+$19.71 in the
broad and narrow-band images is striking.  This difference is due to
strong, redshifted [OIII] 5007\AA ~ emission.  The [OIII] emission
line nebula of 4C $+$19.71 is extended for over $8''$, or about 74
kpc, and is highly elongated in the north-south direction, with an
axial ratio of about 8:1.  South of component ``a" the nebula is
nearly linear with a flaring at the end.  Note that component ``a" in
Fig. 2 actually appears to be either double, or have a core plus
``jet" morphology.  The position angle between the primary and
secondary (fainter) parts of component ``a" (``a$_{1}$" and
``a$_{2}$" respectively) is approximately 160$^{\rm \circ}$, and it
thus does not match the position angle of the nebula as a whole.
North of component ``a" the nebula is much more diffuse, possibly
curving to the west at the location of component ``b".  The length of
the nebula is comparable to the separation of the radio lobes mapped
by Rottgering et al. (1994) and the overall position angle of the
[OIII] emission is the same as the position angle of the radio lobes
on the sky is $176^{\rm \circ}\pm3^{\rm \circ}$.  Eales \& Rawlings
(1996) also note that the [OIII] emission appears extended to the
north in their K-band spectrum.

{}From these data the 4C $+$19.71 nebula has an [OIII] 5007\AA~
emission-line flux measured within a $2''\times10''$ rectangular
aperture of 1.54$\times10^{-18}$ W m$^{-2}$, corresponding to a
5007\AA~ line luminosity of about 3$\times10^{37}$ W.  Eales \&
Rawlings (1996) measure an [OIII] 5007\AA~ line flux of
$\sim10^{-18}$W m$^{-2}$ through a $3.1''\times3.1''$ slit, and
$\sim2\times10^{-18}$ W m$^{-2}$ through a $3.1''\times12.4''$ wide
slit, oriented at a position angle of $162^{\rm \circ}$.  By
comparing the line flux to the total emission measured through the
K-band filter, we estimate that the 5007\AA~ line alone comprises
approximately 34\% of the broad K-band flux.  The rest
frame equivalent width of the 5007\AA~ line, when averaged over the
entire nebula, is $560\pm58$\AA~.  Note that the total [OIII] line
contribution to the measured K-band flux is about 45\%.

The nature of the K$\sim19.8$ mag source (G6) located approximately
$2.1''$ northwest of component ``b" is unknown.  It's proximity to
the radio source argues that it may be associated, expecially since
it lies along what appears to be a bend in the emission-line nebula
beyond component ``b".  However, G6 has no discernible excess of flux
in the narrow band filter, indicating that there is no strong [OIII]
emission from this source.  Given its lack of a narrow-band color
excess and a K$=19.78\pm0.05$ mag, we calculate a line flux limit of
about $3\times10^{-20}$ W m$^{-2}$ from G6.  If G6 is at the redshift
of 4C $+$19.71, this corresponds to a limit on its [OIII] line
luminosity of about $7\times10^{35}$ W, or about 2\% of that measured
for the radio galaxy nebula as a whole.

In Fig. 3 we present a plot of the difference between the broad and
narrow-band magnitudes (the color ``excess") versus the broad-band
magnitude for the objects in the 4C $+$19.71 field.  All of the
sources identified in Fig. 1 and Table 1 are plotted here, except for
G8 and obj 12, which have been excluded because the uncertainties in
their narrow-band magnitudes are larger than 50\%.  The large
narrow-band color excess of 4C $+$19.71 is obvious in Fig. 3.  We
have indicated two points for 4C $+$19.71 in Fig. 3 - components ``a"
and ``b" as defined above.  The different locations of the two points
in Fig.  3 reflect the variation of the [OIII] 5007\AA ~
emission-line equivalent width along the extent of the 4C $+$19.71
nebula, in the sense that the equivalent width of the nebula is
larger away from the galaxy nucleus.  This simply reflects the fact
that there is very little continuum light in the galaxy (at least to
the limits of this image) at large radii along the nebular axis.

Besides 4C $+$19.71 itself, there are three sources with an apparent
narrow-band excess at the $3\sigma$ level or above.  These are the bright
resolved sources labelled G1 and G2, and the bright, apparently stellar
source labelled star B.  Sources G1 and G2 have narrow-band excesses at
the $3-3.5\sigma$ level, while star B has an excess at closer to the
$5\sigma$ level.  Object 10 has a large narrow-band excess (0.41 mag), but
the uncertainties are large ($\pm0.23$ mag).  The other nine sources
plotted in Fig. 3 scatter about the K-NB$=0.0$ mag line, with a total
dispersion of about 0.2 mag.  An excess of 0.2 mag corresponds to an
[OIII] 5007\AA~ rest frame equivalent width of about 13\AA~ at the
redshift of 4C $+$19.71.

\section{Discussion}

Since 4C $+$19.71 is one of only three galaxies at $z>3$ to be imaged
in an emission line, it is instructive to place the 4C $+$19.71
nebula within the context of those seen around other powerful radio
galaxies.  The two largest sets of radio galaxy narrow-band imaging
data are those compiled by Baum et al. (1988) and McCarthy, Spinrad
\& van Breugel (1995).  The low redshift ($0.003<z<0.48$) 3CR sources
imaged by Baum et al. (1988) have visual emission-line luminosities
of $2.2\times10^{32}-1.3\times10^{36}$ W (median
L$\sim3\times10^{34}$ W) and sizes of $1-96$kpc (median diameter,
d$\sim10$ kpc).  The intermediate redshift ($0.058<z<1.847$) 3CR
nebulae imaged by McCarthy, Spinrad \& van Breugel (1995) have visual
emission-line luminosities of $2\times10^{33}-2\times10^{37}$ W
(median L$\sim6\times10^{35}$ W), corrected to [OIII] 5007\AA~ using
emission-line flux ratios of [OIII]/H$\alpha+$[NII]=1,
[OIII]/Ly$\alpha$=0.3, and [OIII]/[OII]=3, and sizes of $1-213$ kpc
(median d$\sim30$ kpc).  Five systems in the McCarthy, Spinrad \& van
Breugel sample (about 15\%) at redshifts of $z>0.5$ have nebulae
larger than 100 kpc, as measured down to a surface brightness level
of $10^{-20}$ W m$^{-2}$ arcsec$^{-2}$.  There are nebula found in
both data sets with large, elongated features (e.g. 3C 227, 3C 458,
3C 435A), yet none are dominated by a single, high surface
brightness, high axial ratio nebulosity as seen in 4C $+$19.71.

Although the the Baum et al. (1988) and McCarthy et al. (1995)
samples are the largest existing narrow-band imaging data sets, most
of the objects are at redshifts well below $z\sim2$.  From a sample
of steep spectrum 4C radio sources, Chambers et al. (1996a,b) image
five galaxies with $2.348<z<2.905$ through narrow-band filters
matched to the Ly$\alpha$ line.  The Ly$\alpha$ nebulae of these
galaxies range from about $25-90$ kpc in extent, and are generally
aligned along the radio axis.

At $z>3$ there are two systems previously known to possess large,
emission-line nebulae.  A large ($\sim60$ kpc), apparently rotating
Ly$\alpha$ halo has been imaged around the $z=3.57$ radio source 4C
03.24 by van Ojik et al. (1997).  The Ly$\alpha$ luminosity of the
nebula in 4C 03.24 is about $6\times10^{37}$ W.  Graham et al. (1994)
imaged the [OIII]$+$H$\beta$ nebula around the $z=3.8$ radio galaxy
4C $+$41.17 by taking the difference between images in a standard
K-band and a K$_{s}$ ($2.0-2.3\mu$m) filter.  These authors show that
the nebula has an extent of about $15\times40$ kpc, and that it is
oriented along the radio axis.  However, unlike 4C $+$19.71, the 4C
$+$41.17 nebula is wider, and appears to extend beyond the main radio
components.  Chambers, Miley \& van Breugel (1990) showed that 4C
$+$41.17 also has a large Ly$\alpha$ nebula which is similarly
extended along the radio axis.

\subsection{Relation of [OIII] and Radio Properties of 4C $+$19.71}

4C $+$19.71 is a double-lobed, FRII radio source with a lobe
separation of 8.1$''$ at a position angle of $176^{\rm \circ}$.
There is no obvious central radio core.  The monochromatic power at a
rest frequency of 6.76 GHz is approximately $10^{28}$ W Hz$^{-1}$.
The extent of the [OIII] nebula matches the separation of the radio
lobes in 4C $+$19.71, and the overall position angle of the nebula is
similar to the position angle of the radio lobes.  Although the
absolute positioning of the infrared images is uncertain, and the
radio map has no strong core, an R-band image of the field covering
10 arcminutes and including six HST guide stars as well as ``star A"
indicates that to within an uncertainty of $\sim2''$, the lobes are
centered on component ``a$_{1}$".  If ``a$_{1}$" is the nucleus of
the radio galaxy, and also the centroid of the two radio lobes, then
the radio lobes are located on the [OIII] image as depicted in Fig.
2.  This positioning suggests that a narrow, slightly curving
``corridor" of line-emitting gas fills the regions between the galaxy
nucleus and the radio lobes.  This gas does not extend beyond the
radio lobes to a limiting surface brightness of about
2$\times10^{-20}$ W m$^{-2}$ arcsec$^{-2}$.  Note that the position
angle between the primary and secondary parts of component ``a" is
not the same as that of the nebula as a whole or the radio lobe
axis.  Thus if ``a$_{2}$" represents a hotspot in the emission line
gas caused by the interaction of an inner jet with the interstellar
medium of 4C $+$19.71, a precession of the jet, or a bend caused by
the interaction of the jet with an overdense region, may explain the
difference in the position angle between the inner and the outer
nebula.

The emission-line$-$to$-$radio luminosity ratio in 4C $+$19.71 is
entirely consistent with the relation found for other FRII radio
sources over a large range in redshift.  In Fig. 2 of McCarthy (1993)
the luminosity in the [OII] 3727\AA ~line is plotted against the
monochromatic radio power at an observed frequency of 1400 MHz.  A
clear correlation exists over five orders of magnitude in radio and
emission-line power.  If the [OIII] 5007/[OII] 3727 line flux ratio
in 4C $+$19.71 is 3.0, the [OII] luminosity is about $10^{37}$ W.
Similarly the radio spectral index of $\alpha = -1.24$ (Rottgering et
al.  1994), where f$_{\nu} \propto \nu^{\alpha}$, implies a 1400 MHz
flux density of $\sim3\times 10^{-27}$ W m$^{-2}$ Hz$^{-1}$, and thus
a monochromatic power at an emitted frequency of 6.43 GHz of about
1.4$\times10^{28}$ W Hz$^{-1}$. For this radio power, the predicted
[OII] luminosity of 4C $+$19.71 is slightly larger than the average
value, yet well within the scatter in the McCarthy plot.  If $\alpha
= -1.24$ out to an observed frequency of about 300 MHz, the
monochromatic power at an emitted frequency of 1400 MHz is closer to
$10^{29}$ W Hz$^{-1}$, thus placing 4C $+$19.71 near the center of
the emission-line$-$radio power correlation.  Thus both the
morphology and the overall luminosity of the optical emission-line
gas and radio plasma in 4C $+$19.71 suggest a causal link between the
relativistic electrons and the 10$^{4}$K ionized gas.

There are a number of possible explanations for the correlation of
the optical emission-line gas and the radio plasma in 4C $+$19.71.
This gas could be photoionized by the active nucleus, or ionized by
shocks associated with the radio jets responsible for supplying the
radio lobes with energetic electrons.  The emission-line gas may form
a shell or channel around the radio jet as has been seen in the
nearby Seyfert galaxy NGC 1068 (Gallimore et al. 1996, Capetti,
Macchetto \& Lattanzi 1997).  Alternatively, the expanding jets could
trigger star formation in the host galaxy, and these hot stars could
ionize the surrounding medium.  This would indicate a more indirect
connection between the radio emission and the optical line-emitting
gas.  The strong alignment in position angle between the radio lobes
and the [OIII] gas, and the similarities in their overall dimensions
could be used to argue for any or all of these possibilities.  The
560\AA ~rest frame equivalent width of the [OIII] line is much larger
than what is seen in starburst galaxies, yet is comparable to the
largest values seen in powerful radio galaxies, being a factor of
$\sim2$ larger than the ``typical" value tabulated by McCarthy
(1993).  Thus it is unlikely that the bulk of the [OIII] emission is
produced by young stars.  Longslit, near infrared spectroscopy of the
$2.3\mu$m spectral region at high spectral resolution could provide a
more concrete assessment of the source of ionizing photons through a
measurement of the the linewidth and flux ratio of the [OIII] 5007\AA
~and H$\beta$ emission lines.  In the spectrum presented by Eales \&
Rawlings (1996) the H$\beta$ emission-line is undetected, suggesting
that the [OIII] 5007/H$\beta$ line flux ratio is greater than 7-10,
consistent with a hard source of ionizing photons (Veilleux \&
Osterbrock 1987).

\subsection{Dust and Gas in 4C $+$19.71}
 
\subsubsection{Dust}

By comparing the [OIII] 5007\AA~ emission-line flux reported here
with the flux of another emission-line whose ratio to [OIII] is
known, we can estimate the amount of reddening towards the ionized
gas in 4C $+$19.71.  As pointed out by Dey, Spinrad \& Dickinson
(1995), the fact that there are only three known dusty, high redshift
galaxies implies that these objects are either intrinsically very
rare, or they have been systematically missed in surveys due to
observational biases.  However, since the presence of significant
amounts of dust in a high redshift galaxy requires the dust to have
been formed at an even earlier epoch, dusty systems at $z>2$ may
imply very early star formation episodes.

Recombination lines are best suited to the task of determining
reddening toward the emission-line gas, but pairs of these have not
yet been measured in 4C $+$19.71.  However, a total Ly$\alpha$ flux
has been measured in 4C $+$19.71 to be about $1.1\times10^{-18}$ W
m$^{-2}$ (H. Spinrad \& M. Dickinson, private communication).  Thus
the measured [OIII] 5007\AA to Ly$\alpha$ line flux ratio, averaged
over the entire 4C $+$19.71 nebula, is $\sim1.4$.  We can estimate
the amount of reddening toward 4C $+$19.71, by assuming the intrinsic
spectrum is similar to the ``average" radio galaxy spectrum compiled
by McCarthy (1993).  Since powerful radio galaxies typically have
[OIII] 5007/Ly$\alpha$ line flux ratios of about 0.3 (McCarthy 1993),
the increased ratio measured for 4C $+$19.71 may indicate the
presence of dust.  To account for the increase in the [OIII]
5007/Ly$\alpha$ line flux ratio seen here would require a visual
extinction of only A$_{\rm V}\sim1$ mag.  With an $l^{\rm II}=75^{\rm
\circ}$ and a $b^{\rm II}=-30^{\rm \circ}$, the Galactic color excess
towards 4C $+$19.71 is E(B-V)$\sim0.04-0.05$ mag (Burstein \& Heiles
1982), implying an A$_{\rm V}\sim0.12-0.16$ mag, negligible in
comparison to the amount of extinction required if the intrinsic
[OIII] 5007/Ly$\alpha$ line flux ratio is 0.3.  An A$_{\rm V}\sim1$
mag is not very large, yet it implies a significant mass of
associated HI gas, if the dust is spread out over the entire nebula.
If A$_{\rm V}=5.3$N$_{\rm H}/10^{22}$ mag (Bohlin, Savage \& Drake
1978), the HI column density toward the emission-line gas in 4C
$+$19.71 is about $2\times10^{21}$ cm$^{-2}$.  If this gas is spread
out over a projected surface area of $8''\times1''$ or 691 kpc$^{2}$,
the implied HI gas mass is about $10^{10}$M$_{\odot}$.  An
atomic$-$to$-$molecular gas ratio similar to the Galaxy thus implies
a similarly large mass of H$_{2}$ (Young \& Scoville 1991).  Of
course, if Ly$\alpha$ is resonantly scattered throughout the nebula,
a very small amount of dust could be responsible for the depleted
line flux, in turn implying much less HI, and by inference, much less
molecular gas in 4C $+$19.71.

Under dusty conditions we would expect the nebula as seen in
Ly$\alpha$ to have a different morphology and perhaps be much smaller
in extent than the nebula as seen in [OIII]. However, images of 4C
$+$19.71 in the Ly$\alpha$ line (L. Maxfield \& M. Dickinson, private
communication) suggest the nebula has a projected size which is at
least as large as the [OIII] nebula imaged here.  Dust which is not
uniformly distributed over the face of the galaxy, perhaps in a dusty
disk or torus around the nucleus, could serve to remove much of the
rest frame UV from our line of sight yet still allow ionizing photons
to escape along the radio jet axis, as in the standard AGN
unification model (e.g.  Antonucci 1993).  Alternatively the
``average" extinction we have calculated may be dominated by small,
dense clumps of dust and gas scattered throughout the nebula.

Although the average visual extinction required to bring the observed
4C $+$19.71 [OIII] 5007/Ly$\alpha$ line flux ratio in line with most
other powerful radio galaxies is small, it is predicated upon the
assumption that the nebula is photoionized.  In general this is a
good assumption, since photoionization models can reproduce the
emission-line flux ratios measured in many radio galaxies (McCarthy
1993).  However, if there are significant dynamical ionization
processes such as shocks, the intrinsic [OIII] 5007/Ly$\alpha$ flux
ratio can be as low as 0.04 (Dopita \& Sutherland 1995).  In this
case the required visual extinction to the 4C $+$19.71 nebula would
be A$_{\rm V}>2$ mag, using the measured [OIII] 5007\AA ~ and
Ly$\alpha$ line fluxes.

\subsubsection{Ionized Gas}

The size of the 4C $+$19.71 [OIII] nebula implies a distribution of
potentially metal-rich gas on galactic scales.  To determine the mass
of ionized gas around 4C $+$19.71 we must know either the gas
density, or the volume and volume filling factor of the nebula.  For
a collisionally excited transition such as the [OIII] 5007\AA ~ line,
the cooling rate per unit volume in the low density limit is
N$_{2}$A$_{21}$h$\nu_{21}=$n$_{e}$N$_{1}$q$_{12}$h$\nu_{21}$
(Osterbrock 1974), where N$_{2}$ is the number density of O$^{++}$
atoms in the excited state, N$_{1}$ is the number density of O$^{++}$
atoms in the lower state, A$_{21}$ is the radiative transition
probability, and q$_{12}$ is the collisonal excitation rate.  For the
[OIII] 5007\AA ~line, A$_{21}$h$\nu_{21}=7.5\times10^{-14}$ erg
s$^{-1}$ and q$_{12}=4.1\times10^{-9}$ cm$^{-3}$s, for T$=10^{4}$K
(Osterbrock).  The mass of O$^{++}$ can then be written as,

\begin{equation} 
{\rm M}_{\rm OIII}=8.7\times10^{7}{\rm
L}_{44}^{5007}{\rm n}_{e}^{-1} {\rm M}_{\odot} 
\end{equation}

\noindent
where L$_{44}^{5007}$ is the luminosity in the 5007\AA ~ line in 
units of $10^{44}$ erg s$^{-1}$.  Although we do not know the ionization 
state of the gas from a single measurement of [OIII], if 
the ratio of O$^{++}$/O in the nebula is about 0.3 (as it is in Cygnus A - 
Osterbrock), 
the total mass of ionized hydrogen is, 

\begin{equation}
{\rm M}=1.4\times10^{11}({\rm n}_{e}{\rm Z})^{-1} {\rm M}_{\odot},
\end{equation}

\noindent
where Z is the O/H ratio of the nebular gas in solar 
units.  Thus if n$_{e}=10^{2}$ cm$^{-3}$ and Z$=1$, the total
mass of ionized gas in the 4C $+$19.71 is about $10^{9}$ M$_{\odot}$.
This is comparable to the most massive nebulae seen around low 
redshift radio galaxies in the light of visual emission-lines 
such as [OIII] 5007\AA ~and H$\alpha +$[NII] (Baum \& Heckman 1989).

By combining these mass estimates with those made from a
recombination line such as Ly$\alpha$ or H$\alpha$, we can solve for
the O/H ratio, given assumptions about the volume and the volume
filling factor of the gas.  For pure case B recombination, the mass
of ionized gas in the nebula as determined from the luminosity in the
Ly$\alpha$ line is
M$=1.3\times10^{8}$(f$_{5}$V$_{68}$L$_{44}^{Ly\alpha})^{1/2}$M$_{\odot}$
or M$=2.1\times10^{10}$L$_{44}^{Ly\alpha}$n$_{e}^{-1}$ M$_{\odot}$,
where f$_{5}$ is the volume filling factor in units of $10^{-5}$,
V$_{68}$ is the volume of the nebula in units of $10^{68}$ cm$^{3}$,
and L$_{44}^{Ly\alpha}$ is the Ly$\alpha$ line luminosity in units of
$10^{44}$ erg s$^{-1}$.  From measurements of the gas density in the
optical line-emitting nebula of radio galaxies at low redshift,
typical volume filling factors of $10^{-4}-10^{-5}$ have been derived
(Heckman, van Breugel \& Miley 1984).  Assuming cylindrical symmetry,
we estimate the volume of the nebula in 4C $+$19.71 to be about
$1.5\times10^{68}$ cm$^{3}$.  Thus, if f$_{5}=1$, V$_{68}=1.5$, and
L$_{44}^{Ly\alpha}=2.1$ (H. Spinrad, private communication),
M$=2.3\times10^{8}$ M$_{\odot}$, and n$_{e}=190$ cm$^{-3}$.  This
assumes that the Ly$\alpha$ emitting nebula is of similar volume as
the [OIII] nebula mapped here, and that A$_{\rm V}=0$ mag.  Note that
if A$_{\rm V}=1$ mag for the nebula and f$_{5}=1$, the implied mass
is closer to $10^{9}$ M$_{\odot}$ and the electron density is about
675 cm$^{-3}$.

Equating the masses derived from both the [OIII] and the Ly$\alpha$
lines suggests that Z$\sim3$ for the nebular gas if A$_{\rm V}=0$
mag.  A galaxy at $z=3.6$ might be expected to be in the early stages
of its evolution and thus have low overall metallicity.  Indeed, many
damped Ly$\alpha$ absorption line systems at high redshifts appear to
have very low metallicities ranging from $10^{-3}-10^{-2}$ of the
solar value (e.g. Lu et al. 1996), consistent with hierarchical
models of structure formation (e.g. Rauch, Haehnelt \& Steinmetz
1997, Hellsten et al. 1997).  However, AGN emission lines typically
imply solar or above metallicities over a large range in redshift.
In high redshift quasars, this has been taken to imply early, and
vigorous star formation pre-dating the AGN phase (Hamann \& Ferland
1992,1993).  If the [OIII] nebular gas in 4C $+$19.71 is metal-rich,
it must have been processed through stars before the observed epoch.
At $z=3.6$, the maximum age of a stellar population is about 2 Gyr.
This maximum age is closer to $5\times10^{8}$ yrs if the formation
redshift is $z\sim5$.  By comparison, values of Z$\sim$3 are
reproduced in the galactic chemical evolution models of Matteucci \&
Padovani (1993) for massive galaxies at ages of only about 0.3 Gyr
for Salpeter (or flatter) IMF slopes.  Flatter IMF spectral indices
can generate O/H ratios of a few times solar after only 0.1 Gyr.
While O/H ratios above solar are easily explained in terms of an
early starburst for 4C $+$19.71, an A$_{\rm V}=1$ mag for the nebula
would imply an O/H ratio of about 0.7 solar, for values of the
filling factor and nebular volume given above.  Since extinctions
much larger than A$_{\rm V}=1$ mag probably do not apply to the 4C
$+$19.71 nebula overall, unless the intrinsic [OIII]
5007\AA$-$to$-$Ly$\alpha$ line flux ratio is significantly different
than what is seen in other powerful radio galaxies, it is unlikely
that the emission-line gas has a metallicity far below solar.
However, another uncertainty in this calculation is the temperature
of the line-emitting gas.  Lower metallicities imply hotter
temperatures, stronger [OIII] emission, and lower oxygen masses.  If
T$\sim2\times10^{4}$K, the oxygen mass is about 36\% of the value
estimated above, and the O/H ratio closer to solar for an A$_{\rm
V}=0$ mag.  Thus, while early enrichment by a previous ($z>3.6$)
episode of star formation is consistent with the data, the derivation
of the metallicity of the gas is highly uncertain and values of the
O/H ratio from a few tenths to a few times solar are possible.

\subsubsection{Pressure Balance in the 4C $+$19.71 Radio Lobes}

The radio properties of 4C $+$19.71 (under the assumption of pressure
balance between the radio plasma and the gas), can also be used to
estimate the mass of the nebula.  Following Miley (1981), we can
write the magnetic field strength, under energy equipartition, as
B$\sim1.5\times10^{-3}$[F$\rm _o$/$\theta^{2}$S]$^{2/7}$ G, where
F$\rm _o$ is the measured flux density of the source in Jy,
$\theta^{2}$ is the surface area of the lobe in arcsec$^{2}$, and S
is the path length through the source in kpc.  If the radio lobes
have a spectral index of $\alpha=-1.24$, the volume filling factor of
the synchrotron emitting plasma is unity, and the radio lobes occupy
spheres of 1$''$  (about 9 kpc) diameter, each with a flux density of
0.15 Jy at 1.465 GHz, the minimum energy condition implies a magnetic
field of $\sim5\times 10^{-4}$ G.  If the radio lobes are confined by
static, thermal pressure from a gas at $10^{4}$K, a gas density of
more than $6\times10^{-21}$ gm cm$^{-3}$ is implied.  However, if the
lobes are instead confined by ram pressure and we use the axial ratio
of the emission line gas (approximately 5:1 from core to lobe) and a
typical lobe advance speed of v$\sim0.01$c (Readhead et al. 1996) we
can estimate a lateral velocity of v$\sim 0.002$c.  The implied gas
density is then greater than about $2\times10^{-25}$ gm cm$^{-3}$.
If the volume filling factor is $10^{-5}$, the densities derived from
the pressure equilibrium arguments imply gas masses of about
$5\times10^{9}$ M$_{\odot}$ and $4\times10^{6}$ M$_{\odot}$,
respectively, if the densities are representative of the entire
nebula.

\subsection{Continuum Luminosity of 4C $+$19.71}

In a $2''\times10''$ beam, the [OIII] 5007\AA ~line contributes
approximately 34\% of the total flux in the K band, and the observed
$2.2\mu$m continuum of 4C $+$19.71 therefore has K$\sim19.6$ mag, if
the [OIII] 4959\AA ~line has 1/3 the flux of the 5007\AA ~line.  We
have not subtracted a contribution from H$\beta$ to derive the
continuum magnitude, but this is likely to be relatively small,
perhaps 10\% the flux of the 5007\AA ~line (Eales \& Rawlings 1996).
In order to compare the continuum magnitude to that expected from
``normal" galaxies at $z\sim3.6$, we employ the luminosity function
of Mobasher, et al. (1993), such that an E/S0 galaxy has a
characteristic absolute magnitude, M$_{B}^\ast = -20.24$ mag,
following Schecter (1976).  Combining this characteristic absolute
magnitude with the spectral energy distributions of Coleman, Wu \&
Weedman (1980), we estimate that the apparent K-band magnitude of an
L$^{\ast}$ elliptical galaxy at $z\sim3.6$ is K$\sim23.5$ mag.  Thus,
if the rest frame visual continuum light from 4C $+$19.71 is produced
by stars, it implies a 40L$^{\ast}$ host galaxy.  This is extremely
luminous, yet the host galaxies of some radio loud quasars (Heckman
et al. 1991; Lehnert et al.  1992; Armus et al. 1997) as well as
radio galaxies (Eales \& Rawlings 1993, 1996; Evans et al. 1996) at
redshifts of $z\sim2-3$ appear to have rest frame visual luminosities
in this range.  Although significant fading (by 2-4 magnitudes in the
rest frame visual) would be required for these systems to evolve into
galaxies resembling even the brightest cluster ellipticals at the
present epoch, this amount of fading can be accomodated by the
postburst models of Chambers \& Charlot (1990) and Charlot \& Bruzual
(1991).  Of course, if much of the continuum emission at these
wavelengths in 4C $+$19.71 is scattered AGN light, the host galaxy
may be significantly fainter than K$\sim19.6$ mag, and the amount of
fading required would be correspondingly lower.

\subsection{Clustering}

Although the goal of this paper is not to search for young clusters
or ``protogalaxies", it is worthwhile to explore the possibility of
clustering around 4C $+$19.71.  At $z=3.594$ an L$^{\ast}$ spiral
galaxy has K$\sim23.7$ mag.  Since the $3\sigma$ point source
detection limit our K-band image is K$\sim 22.5$ mag, objects at the
redshift of 4C $+$19.71 would be difficult to detect in the continuum
unless they were at least one magnitude brighter than an L$^{\ast}$
galaxy.  However, strong sources of line emission at specific
redshifts ($z\sim3.6$ for [OIII] or $z\sim2.5$ for H$\alpha$) could
be detected through their excess light at $2.3\mu$m as measured with
the narrow band filter.

Besides 4C $+$19.71 itself, there are three sources with an apparent
narrow-band excess at the $3\sigma$ level or above.  The bright
sources G1 (K$\sim16.0$ mag) and G2 (K$\sim15.4$ mag) are well
resolved, and it is likely they are foreground to 4C $+$19.71.
Infrared lines such as Br$\gamma$ at $z\sim0.062$, H$_{2}$ (1-0 S(1))
at $z\sim0.084$, HeI at $z\sim0.118$, or Pa$\alpha$ at $z\sim0.227$
are possibilities for the source of the excess in G1 and G2.  The
third source showing a significant ($\sim5\sigma$) narrow-band flux
excess is star B.  This source is intriguing since it is unresolved
in the K-band image, but slightly elongated in the narrow-band
$2.3\mu$m image.  The true nature of this source is unknown, but its
brightness (K$\sim16.7$ mag) argues against it being at a redshift of
$z=3.6$.  The other ten sources plotted in Fig. 3 scatter about the
K-NB$=0.0$ mag line, with a total dispersion of about 0.2 mag, except
for obj 10 which has a large uncertainty in its narrow-band excess.
Since an excess of 0.3 mag corresponds to a rest frame equivalent
width of about 10\AA ~ at the redshift of 4C $+$19.71, none of the
remaining sources can be emission-line objects at the redshift of 4C
$+$19.71 with [OIII] emission-line equivalent widths of more than
about 2\% of that measured for 4C $+$19.71.  A source present at the
$3\sigma$ level in both the K-band and the narrow-band 2.3$\mu$m
images would have an implied [OIII] line flux of
$\sim2\times10^{-20}$ W m$^{-2}$.  The limiting line flux for obj 12
(the faintest narrow-band source in the central part of Fig. 1) is
approximately $10^{-20}$ W m$^{-2}$. If [OIII]/H$\alpha$ $\sim1$,
then this flux limit corresponds to an H$\alpha$ luminosity of about
$2\times10^{35}$ W, and a star formation rate limit of about $17$
M$_{\rm \odot}$ yr$^{-1}$ (Kennicutt 1983).  The co-moving volume
covered by our image is 212 Mpc$^{3}$ at $z=3.6$ (for H$_{\rm o}=75$
km s$^{-1}$ Mpc$^{-1}$ and q$_{\rm o}=0$).  Thus the space density of
objects with K$<22.5$ mag and [OIII] 5007\AA~ emission-line
equivalent widths of more than about 10\AA~ in the vicinity of 4C
$+$19.71 is less than 5$\times10^{-3}$Mpc$^{-3}$.  Deeper images, at
K and in the $2.3\mu$m filter, would be required to place limits on
L$^{\ast}$ or moderately star-forming galaxies around 4C $+$19.71

\section{Summary}

Using a broad-band K and narrow-band $2.3\mu$m filter, we have
observed the $z=3.594$ radio galaxy 4C $+$19.71 with the Near
Infrared Camera on the W.M. Keck Telescope.  These imaging data have
allowed us to determine the following properties of the emission-line
nebula in this high redshift, powerful radio galaxy:

\noindent
(1)  The [OIII] nebula has an extent of about $74\times9$ kpc, and is
very narrow with some flaring and bending at the ends.  The total
luminosity of the nebula is L$_{5007}\sim3\times10^{37}$ W.  The
fraction of the light that is contributed by the 5007\AA ~ line to
the total K-band flux is about 34\%.

\noindent
(2)  The length of the [OIII] nebula is nearly identical to the
separation of the two radio lobes mapped at 1465 MHz by Rottgering,
et al. (1994), and the position angle of the nebula is similar to
that of the radio emission.  4C $+$19.71 falls on the emission-line
vs. radio power correlation found for other powerful radio galaxies
at low and high redshifts.  Taken together, these facts suggest a
direct link between the relativistic electrons and the 10$^{4}$K gas
in 4C $+$19.71.

\noindent
(3)  The ratio of the [OIII] 5007\AA ~ line luminosity to the total
Ly$\alpha$ line luminosity (H. Spinrad and L. Maxfield, private
communication) is $\sim1.4$, larger by a factor of about $4-5$ than
is typically seen in the spectra of powerful radio galaxies (McCarthy
1993).  This may indicate the presence of dust in 4C $+$19.71, or an
enhancement in the nebular O/H ratio.  The nebula-averaged extinction
required to produce the observed [OIII]$-$to$-$Ly$\alpha$ line flux
ratio is only A$_{\rm V}\sim1$ mag if the intrinsic ratio is equal to
the average value.  If resonant scattering of the Ly$\alpha$ line is
not severe, this extinction implies an HI mass of $10^{9}-10^{10}$
M$_{\odot}$, depending upon the distribution of the obscuring dust.

\noindent
(4)  We derive an ionized gas mass of $2\times10^{8}-10^{9}$
M$_{\odot}$ from the [OIII] and Ly$\alpha$ emission-line
luminosities.  The O/H ratio in the nebula is at
least a few tenths solar, and may be as high as a factor of three
above solar. The latter would argue for a starburst at $z>3.6$ in 4C
$+$19.71.  The gas masses derived by requiring that the radio lobes
be either in thermal pressure or ram pressure equilibrium with the
10$^{4}$K gas are $5\times10^{9}$ M$_{\odot}$ and $4\times10^{6}$
M$_{\odot}$, respectively.

\noindent
(5)  The continuum magnitude of 4C $+$19.71 is K$\sim19.6$ mag.  At
$z=3.594$, this corresponds to a 40L$^{\ast}$ E/S0 galaxy.  Although
the fraction of the continuum light that is non-stellar in origin is
not known, this value for the continuum flux places 4C $+$19.71 along
the K-z relation found for radio loud
quasars and radio galaxies.

\noindent
(6)  There are three objects, besides 4C $+$19.71, which have a flux
excess in the narrow band filter. Two of these are bright
(K$\sim15.4-16.0$ mag) galaxies which are likely to be at low
redshifts.  The nature of the third object is unknown, but its
brightness (K$\sim16.7$ mag) also argues for a similarly low
redshift.  There are no candidate emission-line objects at the
redshift of 4C $+$19.71 having [OIII] rest frame equivalent width of
more than about 2\% of the radio galaxy itself within a co-moving
volume of 212 Mpc$^{3}$.  Thus the space density of objects with
[OIII] emission-line luminosities of $2-3\times10^{35}$ W and rest
frame blue luminosities greater than 3L$^{\ast}$ around 4C $+$19.71
is less than about $5\times10^{-3}$ Mpc$^{-3}$.

\acknowledgments

The W.M. Keck Observatory is operated as a scientific partnership
between the California Institute of Technology and the University of
California.  We thank the entire Keck staff, especially Wendy
Harrison, for making these observations possible.  In addition, we
thank David Hogg, Matt Lehnert, Pat McCarthy, and Tony Readhead for
many helpful discussions.  Hy Spinrad and Leslie Maxfield were kind
enough to make their unpublished visual data available to us, and we
thank them for that.  The comments of an anonymous referee are also
appreciated.  Infrared astrophysics at Caltech is supported by grants
from NASA.  This research has made use of the NASA/IPAC Extragalactic
Database which is operated by the Jet Propulsion Laboratory, Caltech,
under contract with NASA.

\clearpage
 
\begin{center}
{\bf Figure Captions}
\end{center}

\figcaption[fig1.ps]{A mosaic of the K-band images of field around 4C
$+$19.71.  North is up and east is to the left.  The radio galaxy and
all the detected objects in the field are marked as they are
identified in Table 1.  Components ``a" (at 0.0,0.0) and ``b" of the
radio source, as first identified by Eales \& Rawlings (1996), are
also marked.}

\figcaption[fig2.ps]{Small ($20"\times20"$) sub-frames of the K-band
mosaic (left) and narrow band mosaic (right) of the 4C $+$19.71
field.  In both cases, north is up and east is to the left.  The
positions of components ``a" and ``b" are indicated in the K-band
image.  A vertical bar has been drawn within the narrow-band frame
indicating a projected physical scale of 25 kpc at the redshift of 4C
$+$19.71.  The white crosses in the narrow-band frame mark the
locations of the 1465 MHz radio lobes as mapped by Rottgering et al.
(1994), assuming they are centered on component ``a" of 4C $+$19.71.}

\figcaption[fig3.ps]{Color magnitude diagram for the sources in the
4C $+$19.71 field.  The broad-band K minus narrow band color (K-NB)
is plotted against the broad-band K magnitude.  All sources from
Table 1 are included as measured through a $2.0''$ diameter circular
aperture, excluding those with uncertainties in the measured
narrow-band magnitude of more than 50\%.  The radio galaxy, measured
at two locations as defined in the text, has a significant excess in
the narrow band filter.  For reference, the horizontal dotted and
dot-dashed lines correspond to the observed K-NB colors of a
$z=3.594$ source with a rest frame emission-line equivalent width of
50\AA ~and 100\AA, respectively.}

\clearpage

\begin{table}
\center{\bf Objects in the 4C $+$19.71 Field}
\begin{center}
\begin{tabular}{ccc}
Name&K&Position\cr
 &mag&E,N($''$) \cr
\hline
\hline

G2&15.39$\pm0.02$&$+$14.6,$-$12.0 \cr
G1&16.02$\pm0.02$&$+$18.7,$-$12.8 \cr
star A&16.10$\pm0.02$&$+$3.3,$-$11.0 \cr
G3&16.53$\pm0.02$&$+$13.0,$-$16.4 \cr
star B&16.66$\pm0.02$&$+$18.4,$+$8.9 \cr
G4&16.97$\pm0.02$&$-$20.0,$-$8.1 \cr
star C&17.36$\pm0.02$&$-$20.7,$-$14.2 \cr
4C$+$19.71a&20.14$\pm0.05$&0.0,0.0 \cr
4C$+$19.71b&21.41$\pm0.16$&$+$0.3,$+$3.6 \cr 
G5&18.94$\pm0.03$&$+$2.1,$-$7.0 \cr
obj 10&19.35$\pm0.07$&$+$12.7,$-$30.8 \cr
G8&19.37$\pm0.07$&$-$7.5,$+$24.5 \cr    
obj 9&19.70$\pm0.05$&$-$13.3,$+$12.3 \cr
G7&19.76$\pm0.05$&$+$10.5,$+$10.1 \cr  
G6&19.78$\pm0.05$&$-$2.3,$+$4.4 \cr 
obj 11&20.62$\pm0.11$&$-$15.9,$-$5.4 \cr
obj 12&20.65$\pm0.09$&$+$16.5,$+$0.3 \cr

%$+$14.6,$-$12.0&15.39$\pm0.02$&G2 \cr
%$+$18.7,$-$12.8&16.02$\pm0.02$&G1 \cr
%$+$3.3,$-$11.0&16.10$\pm0.02$&star A \cr
%$+$13.0,$-$16.4&16.53$\pm0.02$&G3 \cr
%$+$18.4,$+$8.9&16.66$\pm0.02$&star B \cr
%$-$20.0,$-$8.1&16.97$\pm0.02$&G4 \cr
%$-$20.7,$-$14.2&17.36$\pm0.02$&star C \cr
%0.0,0.0&20.14$\pm0.05$&4C$+$19.71a \cr
%$+$0.3,$+$3.6&21.41$\pm0.16$&4C$+$19.71b \cr
%$+$2.1,$-$7.0&18.94$\pm0.03$&G5 \cr
%$+$12.7,$-$30.8&19.35$\pm0.07$&obj 10 \cr
%$-$7.5,$+$24.5&19.37$\pm0.07$&G8 \cr
%$-$13.3,$+$12.3&19.70$\pm0.05$&obj 9 \cr
%$+$10.5,$+$10.1&19.76$\pm0.05$&G7 \cr
%$-$2.3,$+$4.4&19.78$\pm0.05$&G6 \cr
%$-$15.9,$-$5.4&20.62$\pm0.11$&obj 11 \cr
%$+$16.5,$+$0.3&20.65$\pm0.09$&obj 12 \cr

\end{tabular}
\caption{Near infrared magnitudes of the objects in the 
4C $+$19.71 field.  All values are measured in 2.0$''$ circular 
beams.  
Column 3 is the offset of each object, in arcseconds, from the 
position of the center of 
the radio galaxy (defined as component ``a" in Figs. 1 and 2).
Note that the K band magnitudes of 4C $+$19.71 given in this 
table are as measured through the broad band filter, uncorrected 
for the [OIII] emission-line flux (see text).}

\end{center}
\end{table}


\thebibliography{}

\bibitem{} Armus, L., Neugebauer, G., Lehnert, M.D., \& Matthews, K.
1997, \mnras, 289, 621.

\bibitem{} Baum, S.A., Heckman, T.M., Bridle, A., van Breugel, W., \&
Miley, G.K. 1988, \apjs, 68, 643.

\bibitem{} Baum, S.A. \& Heckman, T.M. 1989, \apj, 336, 681.

\bibitem{} Bohlin, R.C., Savage, B.D., \& Drake, J.F. 1978, \apj, 224, 132.

\bibitem{} Bunker, A.J., Warren, S.J., Hewett, P.C., \& Clemens, D.L.
1995, \mnras, 273, 513.

\bibitem{} Burstein, D., \& Heiles, C. 1982, \aj, 87, 1165.

\bibitem{} Capetti, A., Macchetto, F.D., \& Lattanzi, M.G. 1997,
\apj, 476, L67.

\bibitem{} Chambers, K.C., \& Charlot, S. 1990, \apj, 348, L1.

\bibitem{} Chambers, K.C., Miley, G.K., \& van Breguel, W.J.M. 1987,
\nat, 329, 604.

\bibitem{} Chambers, K.C., Miley, G.K., \& van Breguel, W.J.M. 1988,
\apj, 327, L47.

\bibitem{} Chambers, K.C., Miley, G.K., \& van Breguel, W.J.M. 1990,
\apj, 363, 21.

\bibitem{} Chambers, K.C., Miley, G.K., van Breugel, W.J.M., \&
Huang, J.-S. 1996, \apjs, 106, 215.

\bibitem{} Chambers, K.C., Miley, G.K., van Breugel, W.J.M., Bremer,
M.A.R., Huang, J.-S., \& Trentham, N.A. 1996, \apjs, 106, 247.

\bibitem{} Charlot, S., \& Bruzual A., G. 1991, \apj, 367, 126.

\bibitem{}Coleman, G. D., Wu, C-C. \& Weedman, D. W. 
1980, \apjs, 43, 393

\bibitem{} Dey, A., Spinrad, H., \& Dickinson, M. 1995, \apj, 440, 515.

\bibitem{} Dopita, M.A. \& Sutherland R.S. 1995, \apj, 455, 468.

\bibitem{} Dunlop, J. S. Hughes, D. H., Rawlings, S., Eales, S., \&
Ward, M. J.
 1994, \nat, 370, 347

\bibitem{} Eales, S.A., \& Rawlings, S. 1996, \apj, 460, 68.

\bibitem{} Evans, A.S., Sanders, D.B., Mazzarella, J.M., Solomon,
P.M., Downes, D., \& Radford, S.J.E. 1996, \apj, 457, 658.

\bibitem{} Evans, A.S. 1996 Ph.D. Thesis, University of Hawaii.

\bibitem{} Fanaroff, B.L., \& Riley, F.M. 1974, \mnras, 167, 31P.

\bibitem{} Gallimore, J.F., Baum, S.A., \& O'Dea, C.P. 1996, \apj, 458, 136.

\bibitem{} Golombek, D., Miley G.K. \& Negebauer, G. 1988, \aj, 95, 26.

\bibitem{} Graham, J.R., Matthews, K., Soifer, B.T., Nelson, J.E.,
Harrison, W., Jernigan, J.G., Lin, S., Neugebauer, G., Smith, G., \&
Ziomkowski, C. 1994, \apj, 420, L5.

\bibitem{} Hamann, F. \& Ferland, G. 1992, \apj, 391, L53.

\bibitem{} Hamann, F. \& Ferland, G. 1993, \apj, 418, 11.

\bibitem{} Heckman, T. M., Smith, E. P., Baum, S. A., van Breugel, W.
J. M., Miley, G. K., Illingworth, G. D., Bothun, G. D., \& Balick, B.
1986, \apj, 311, 526

\bibitem{} Heckman, T.M., Lehnert, M.D., Miley, G.K., \& van Breugel,
W. 1991, \apj, 381, 373.

\bibitem{} Hellsten, U., Dave, R., Hernquist, L., Weindberg, D.H. \&
Katz, N. 1997, \apj, submitted.

\bibitem{} Hill, G.J., \& Lilly, S.J. 1991, \apj, 367, 1.

\bibitem{} Ivison, R. J. 1995, \mnras, 275, L33

\bibitem{} Kellerman, K.I., Sramek, R.A., Schmidt, M., Schaffer,
D.B., \& Green, R.F. 1989, \aj, 98, 1195

\bibitem{} Kennicutt, R.C. 1983, \apj, 272, 54.

\bibitem{} Lacy, M., et al. 1994, \mnras, 271, 504.

\bibitem{} Le Fevre, O., Deltorn, J.M., Crampton, D., \& Dickinson M.
1997, \apj in press.

\bibitem{} Lehnert, M.D., Heckman, T.M., Chambers, K.C., \& Miley,
G.K. 1992, \apj,393,68.

\bibitem{} Lu, L., Sargent, W.L.W., Barlow, T.A., Churchill, C.W. \&
Vogt, S. 1996, \apjs, 107, 475.

\bibitem{} McCarthy, P.J., Spinrad, H., Djorgovski, S., Strauss,
M.A., van Breugel, W., \& Liebert, J. 1987, \apj, 319, L39.

\bibitem{} McCarthy, P.J., van Breugel, W., Spinrad, H., \&
Djorgovski, S. 1987, \apj, 321, L29.

\bibitem{} McCarthy, P.J., Spinrad, H., van Breugel, W., Liebert, J.,
Dickinson, M., Djorgovski, S., \& Eisenhardt, P. 1990, \apj, 365,
487.

\bibitem{} McCarthy, P.J. 1993, \araa, 31, 639.

\bibitem{} McCarthy, P.J., Spinrad, H., \& van Breugel, W. 1995,
\apjs, 99, 27.

\bibitem{} Mannucci, F. \& Beckwith, S.V.W. 1995, \apj, 442, 569.

\bibitem{} Matteucci, F., \& Padovani, P. 1993, \apj, 419, 485.

\bibitem{} Mazzarella, J.M., Graham, J.R., Sanders, D.B., \&
Djorgovski, S. 1993, \apj, 409, 170.

\bibitem{} Miley, G.K. 1981, \araa, 18, 165.

\bibitem{} Mirabel, I.F., Sanders, D.B., \& Kazes, I. 1989, \apj,
340, L9.

\bibitem{} Mobasher, B., Sharples, R. M.\& Ellis, R. S.  
1993, \mnras, 263, 560

\bibitem{} Osterbrock, D.E. 1974, Astrophysics of Gaseous Nebula (San
Francisco; Freeman).

\bibitem{} Pascarelle, S.M., Windhorst, R.A., Driver, S.P.,
Ostrander, E.J. \& Keel, W.C. 1996, \apj, 456, L21.

\bibitem{} Rauch, M., Haehnelt, M.G., \& Steinmetz, M. 1997, \apj,
submitted.

\bibitem{} Readhead, A.C.S., Taylor, G.B., Pearson, T.J., \&
Wilkinson, P.N. 1996, \apj, 460, 634.

\bibitem{} Rottgering, H.J.A., Lacy, M., Miley, G.K., Chambers, K.C.,
\& Saunders, R. 1994, \aaps, 108, 79.

\bibitem{} Schechter, P. 1976,\apj, 203, 297

\bibitem{} Spinrad, H., et al. 1993 in Observational Cosmology, ASP
conf. ser. 51, p.585

\bibitem{} Thompson, D., Mannucci, F. \& Beckwith, S.V.W. 1996, \aj,
112, 1794.

\bibitem{} van Ojik, R., Rottgering, H.J.A., Miley, G.K., Bremer,
M.N., Macchetto, F., \& Chambers, K.C. 1994, \aap, 289, 54.

\bibitem{} van Ojik, R., Rottgering, H.J.A., Miley, G.K., \&
Hunstead, R.W. 1997, \aap, 317, 358.

\bibitem{} van Ojik, R., Rottgering, H.J.A., Carilli, C.L., Miley,
G.K., Bremer, M.N., \& Macchetto, F. 1997, \aap, in press.

\bibitem{} Veilleux, S. \& Osterbrock, D.E. 1987, \apjs, 63,295.

\bibitem{} Young, J.S. \& Scoville, N.Z. 1991, \araa, 29, 581.
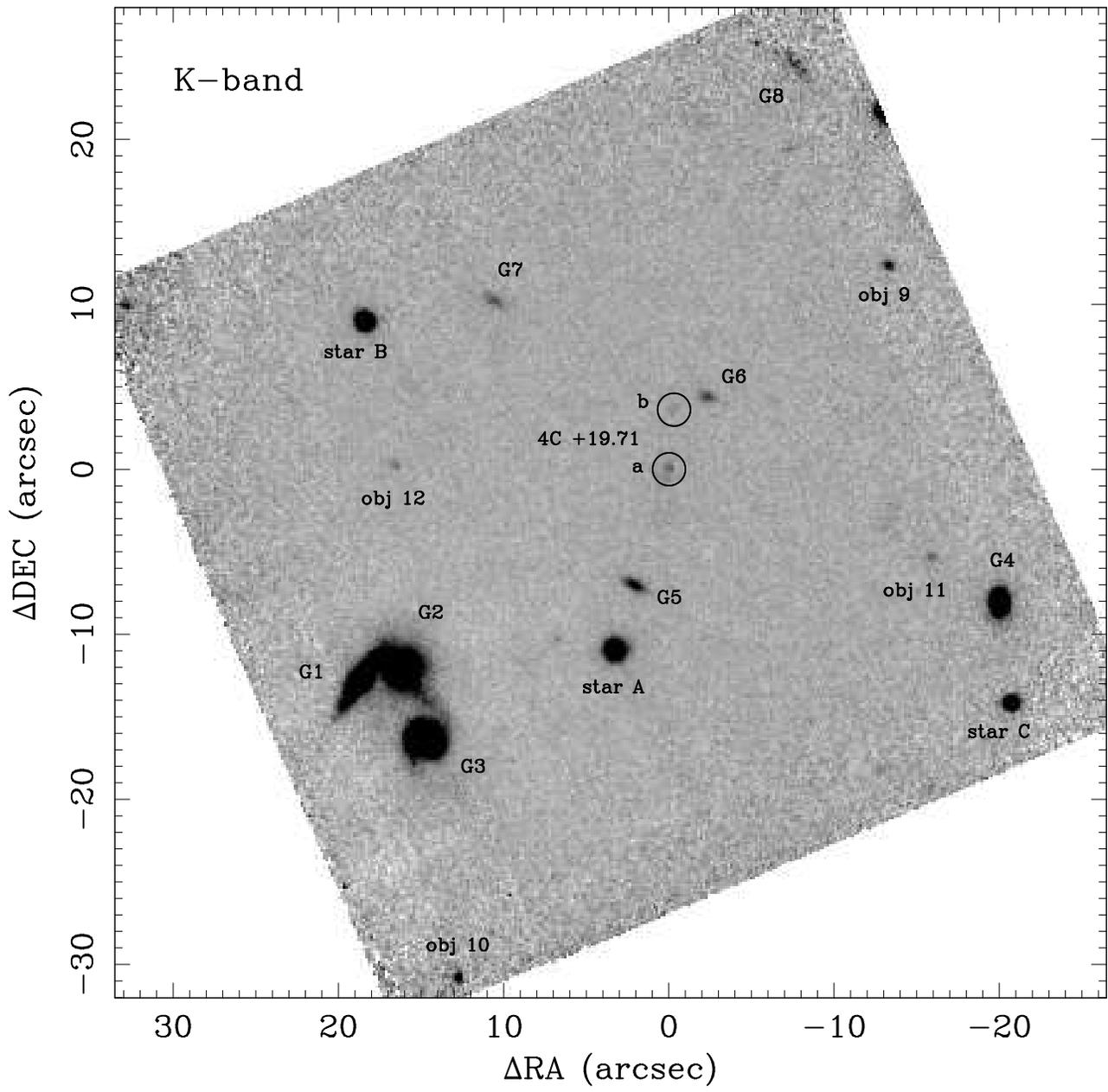
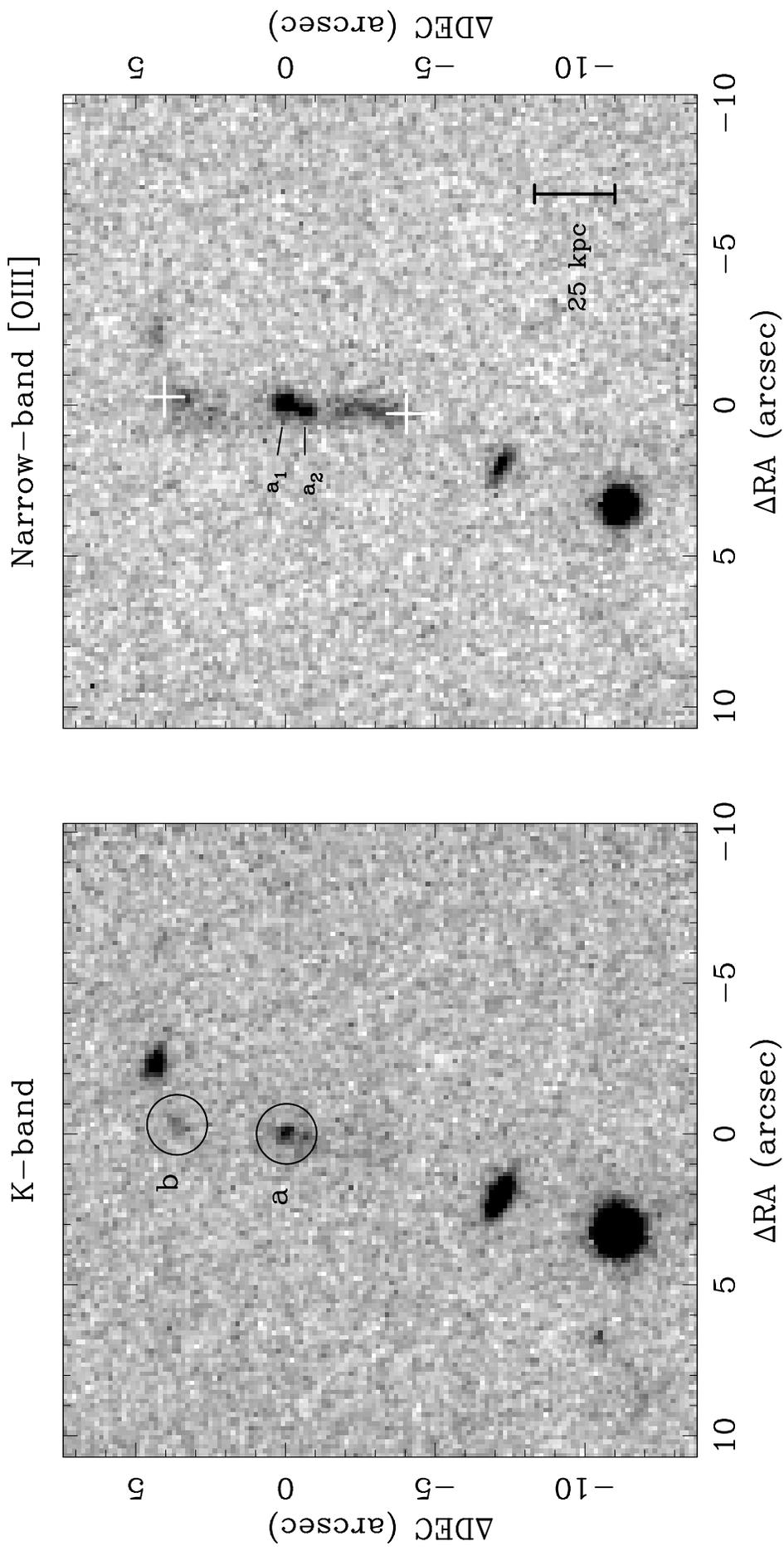
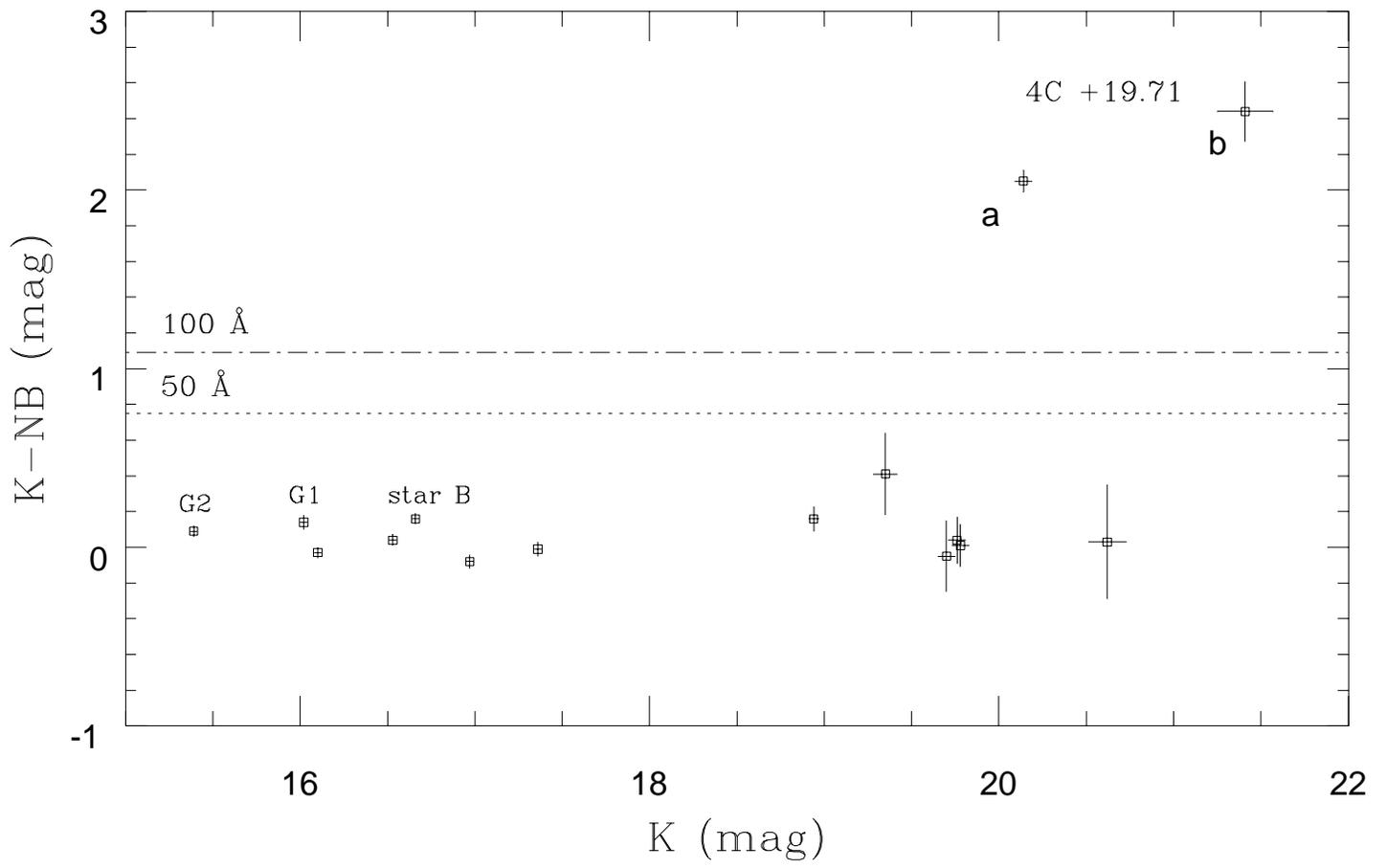
\end{document}